# Acoustically Driven and Modulation Inducible Radiating Elements

Ahmed E. Hassanien*, Michael Breen, Ming-Huang Li & Songbin Gong

University of Illinois at Urbana-Champaign

**Abstract**

The low propagation loss of electromagnetic radiation below 1 MHz offers significant opportunities for low power, long range communication systems to meet growing demands for IoT applications. Especially in the very low frequency (VLF: 3-30 kHz) range and lower, propagation through tens of meters of seawater, hundreds of meters of earth, and hundreds of kilometers of air with only 2-3 dB of loss is possible. However, the fundamental reduction in efficiency as the size of electrical antennas decreases below a wavelength (30 m at 1 MHz) has made portable communication systems in the VLF and low frequency (LF: 30-300 kHz) ranges impractical for decades. A paradigm shift from electrical to piezoelectric antennas utilizing strain-driven currents at resonant wavelengths up to five orders of magnitude smaller than electrical antennas offers the promise for orders of magnitude efficiency improvement over the electrical state-of-the-art. This work demonstrates a lead zirconate titanate transmitter >6000 times more efficient than a comparably sized electrical antenna and capable of bit rates up to 60 bit/s using frequency-shift keying. Detailed analysis of design parameters offers a roadmap for significant future improvement in both radiation efficiency and data rate in the new field of acoustically driven antennas.

**Introduction**

Portable wireless devices have become ubiquitous over the last decade, and with the growth of the internet of things (IoT), demand for small, efficient wireless communication systems continues to accelerate. While the development of wireless systems has kept pace with demand at higher frequencies, progress toward portable low-frequency systems has been stagnant for nearly a century. Compact antennas at the very high frequency (VHF: 30-300 MHz) and ultra high frequency (UHF: 300-3000 MHz) are well developed and suited for transmitting data at high bit rates. However, increased spectral crowding and relatively large propagation loss in those bands make them unsuitable for widespread arrays of remote, low power sensors in rural areas or long-range communication elements over rugged terrain desirable for internet of things (IoT) or defense[1] applications. Compared to VHF and UHF signals, radiation at the ultra-low (ULF: 0.3-3 kHz) and very low frequency (VLF: 3-30 kHz) ranges exhibits relatively low propagation loss, enabling communication underwater up to 20 meters[2] and through hundreds of feet of earth[3]. Additionally, VLF radiation can propagate as ground waves which reflect back and forth between the Earth surface and ionosphere with very low atmospheric attenuation of ~2-3 dB/100 km[4]. However, while the desirable propagation properties ensure continued demand for portable, long-range VLF transmitters, use of VLF antennas has been largely restricted to submarines[5,6] and large base installations[7,8] such as the VLF transmitter in Cutler, Maine.

Despite decades of investigation, compact antennas in the VLF and low frequency (LF, 30 – 300 kHz) bands have remained an unattainable holy-grail considered impractical due to the fundamental tradeoff between antenna efficiency and electrical size. Efficient electrical antennas require operation near electromagnetic resonance, typically restricting the physical size to be larger than one-tenth of a wavelength λ/10. Fundamental analysis on the tradeoff between antenna size and efficiency was first conducted by Wheeler[9] and Chu[10] in the 1940s and extensive subsequent work[11–13] defined the size limit for an efficient electrically small antenna (ESA). Decreasing the size of an ESA below that limit results in a diminished radiation resistance, which leads to a low radiation efficiency as resistive losses begin to dominate[14]. Furthermore, as the size of electrical antennas becomes much smaller than λ, the reactive component of the antenna impedance becomes increasingly large. The small radiation resistance in conjunction with the much larger antenna reactance results in a large impedance mismatch with the driving electronics. Tuning out the large reactance to improve the matching efficiency requires an impedance matching circuit, but for frequencies below 1 MHz, the large size and lossy nature of the matching circuit have made ESAs impractical to implement.

Recently, piezoelectric resonant acoustic antennas have been considered to circumvent the inefficiency of ultra-sub wavelength ESAs required for portable VLF communication. First proposed by Mindlin[15], piezoelectric antennas couple mechanical vibration into electrical radiation. Acoustic waves propagate at velocities $10^5$ times lower than electromagnetic waves, enabling resonant operation for mechanical antennas at frequencies five orders of magnitude lower than similar sized electrical counterparts. Resonant impedances of acoustically driven antennas can be easily matched to driving electronics, removing the need for bulky, inefficient matching circuits and circumventing the Wheeler-Chu limit for VLF antennas. More recently, additional studies on the radiation properties of piezoelectric antennas[16,17] and early prototypes at both VHF[18] and VLF[19] have been demonstrated to show promise as compact antennas with efficiency advantages over ESA.

In this paper, we demonstrate an acoustically driven and modulation inducible radiating element (ADMIRE) using lead zirconate titanate (PZT) as the piezoelectric material which redefines VLF transmitters by circumventing the Chu bandwidth limit and demonstrating novel shaping of near and far-field regions using high-permittivity materials. While the presented antenna efficiency is already more than 6000x that of an equivalently sized ESA, it is still far from the anticipated limits of piezoelectric antennas. A full analysis of the design space for piezoelectric antennas is detailed, paving the way for the subsequent development of compact, high-efficiency piezo-transmitters with the potential for widespread use in low-frequency wireless communication systems.

**Theory**

Acceleration of charges, including dipole moment flipping, results in far-field electromagnetic (EM) radiation with field components that are inversely proportional to the distance traveled away from the radiating element[20]. Using this concept, any element that contains a time-varying dipole moment, such as the acoustically excited piezoelectric materials described in this paper, can be considered a radiating element. Piezoelectric materials lack inversion symmetry within its crystalline structure, resulting in a linear coupling between the electrical and the mechanical domain parameters via the reversible piezoelectric effect. In particular, the direct piezoelectric effect is the electrical polarization produced by an applied mechanical stress. For a time-varying stress, radiation is produced with the time-varying electrical polarization[21].

This concept is further explained in Fig. 1a, where a sinusoidal force, with period $\tau$, is exerted on a piezoelectric material resulting in electric polarization with surface charge density $\sigma_q$ which can be calculated using the piezoelectric constitutive equations as follows[17]:

$$\sigma_q = dT = dC^E S \qquad (1)$$

$$I = \sigma_q A \omega = dC^E S A \omega \qquad (2)$$

where $d$ is the piezoelectric strain constant, $T$ is the applied stress, $C^E$ is the stiffness at constant electric field and $S$ is the resulting strain. The effective dipole current is calculated in (2), where $A$ is the surface area of the accumulated charges and $\omega$ is the angular frequency of the applied stress. The generated magnetic field density in the far-field region due to the dipole current is then formulated as[20]:

$$|B_{far}| = \frac{\sigma_q A}{4\pi\varepsilon_o} \frac{L\omega^2}{c^3 R} \qquad (3)$$

where $L$ is the dipole moment length, $\varepsilon_o$ is the permittivity of the free space, and $c$ is the speed of light. The corresponding far-field electric field $|E_{far}| = c|B_{far}|$.

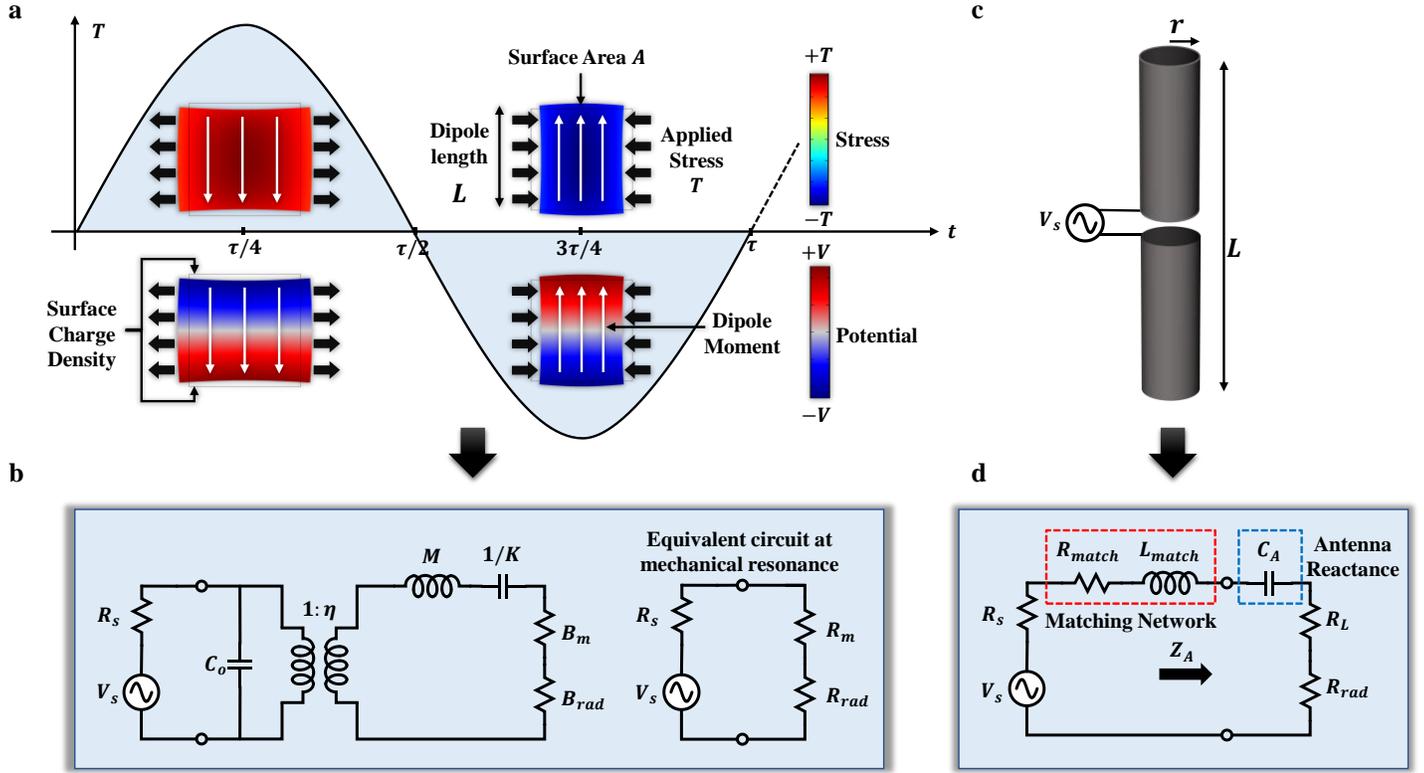

**Fig. 1 Comparison between acoustically driven and modulation inducible radiating elements (ADMIREs) and electrically small antennas (ESAs). a**, ADMIRE-conceptual diagram with a sinusoidal time-varying mechanical stress applied, resulting in time-varying electrical polarization. **b**, Butterworth Van-Dyke equivalent circuit model of the ADMIRE. **c**, ESA representation (infinitesimal dipole). **d**, ESA equivalent circuit model at low frequencies (< 1 MHz) with large antenna reactance dominating the antenna input impedance which requires an impractical matching network.

For comparison, both the ADMIRE and a generic electrically small antenna (ESA), are shown in Fig. 1a and 1c, respectively. Both types of antennas can be modeled in the electrical domain with the equivalent circuit representations shown in Fig. 1b and Fig. 1d, respectively. The ADMIRE is modeled with the Butterworth Van Dyke (BVD) model[22], where at the mechanical resonance the reactive components cancel out ($\omega_r L_m = 1/\omega_r C_m$) and the input impedance of the ADMIRE is $R_m + R_{rad} \ll 1/\omega C_o$. In the model, $R_m$ represents the mechanical losses, $R_{rad}$ is added to represent the radiated power, $C_o$ is the static capacitance between the input electrical terminals, $\omega_r$ is the resonance frequency, $L_m$ is the motional inductance representing the mechanical mass effect, and $C_m$ is the motional capacitance representing the mechanical stiffness effect. As previously explained, ESAs at low frequencies (< 1 MHz) have a large reactive element that requires impractical matching compared with the ADMIRE which is designed to be impedance matched.

It can be shown that the ADMIRE radiation efficiency, defined as the radiated power divided by the input power, is proportional to the piezoelectric material properties and dimensions as follows[17]:

$$\xi_{ADMIRE} \propto d^2 C^E V Q \omega^3 \qquad (4)$$

where $Q$ is the quality factor of the acoustic mode in the material and $V = LA$ is the volume of the ADMIRE. The relative radiation efficiency for similarly sized ADMIRE and ESA can be formulated as[17]:

$$\xi_{rel} = \frac{\xi_{ADMIRE}}{\xi_{ESA}} \propto \frac{d^2 C^E Q \omega}{\sigma_c} \qquad (5)$$

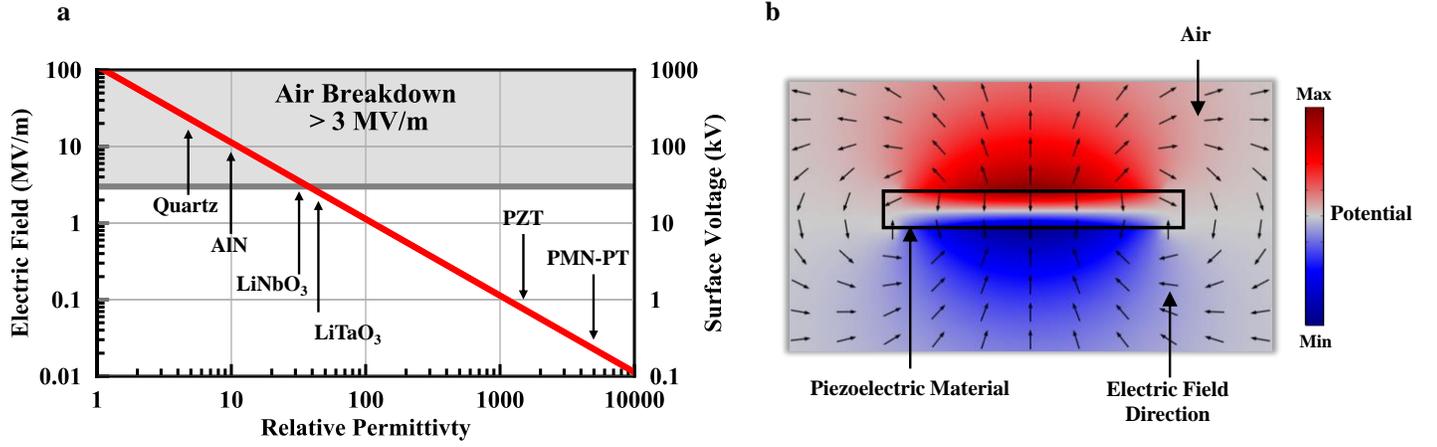

**Fig. 2 Comparison of electrical polarization response for different piezoelectric materials.** All materials are assumed to have a charge density of ±1 mC/m² on the top and bottom surfaces with a thickness (distance between surfaces) of 1 cm. **a**, The electric field and the corresponding surface voltage versus the piezoelectric material relative permittivity. The top gray region is the air breakdown region where the electric field exceeds 3 MV/m. **b**, The voltage distribution due to the electric polarization and the electric field direction represented by the black arrows.

where $\sigma_c$ is the electrical conductivity of the ESA metallic material. From (5), the relative radiation efficiency of the ADMIRE can be increased by selecting a material with larger stiffness, quality factor, and especially the piezoelectric strain constant due to its squared behavior. However, the main advantage of mechanical antennas arises from the mismatch efficiency of ADMIREs compared to ESAs at low frequencies below 1 MHz. The total antenna efficiency is defined as the ratio of the power radiated relative to the maximum available power from the source and is the product of the radiation and mismatch efficiencies. While ADMIREs can be designed to have real resonant impedances that achieve high mismatch efficiency without the need for a matching network at low frequencies, ESAs are well known to exhibit small radiation resistances and large reactive components which result in an enormous mismatch efficiency. To improve total efficiency, ESAs require bulky impedance matching circuits to tune out the reactive component. The relative total antenna efficiency of the ADMIRE, normalized with respect to an impedance matched ESA can be expressed as:

$$\xi_{tot}^{rel} = \frac{\xi_{tot}^{ADMIRE}}{\xi_{tot}^{ESA}} = \frac{R_{rad}^{ADMIRE}}{R_{rad}^{ESA}} \frac{(R_{rad}^{ESA} + R_{loss} + R_{match} + R_s)^2}{(R_{rad}^{ADMIRE} + R_m + R_s)^2} \quad (6)$$

where $R_{rad}^{ADMIRE}$ and $R_{rad}^{ESA}$ are the ADMIRE and the ESA radiation resistances respectively, $R_{loss}$ is the ESA conduction/dielectric losses, $R_{match}$ is the matching resistance resulting from the finite quality factor of the matching inductor, and $R_s$ is the source resistance as shown in Fig. 1. Even with matching networks for the ESAs, typically consisting of low-frequency inductors with quality factors less than a few hundred, the matched impedance seen by the source remains in the kilo-ohms range, resulting in total antenna efficiencies more than 6400x greater in favor of ADMIREs over ESAs.

In addition to the material properties essential for efficient radiation, the relative permittivity of the piezoelectric material bears crucial consideration for reliable antenna operation. As the bound charge densities on the top and bottom surfaces of the ADMIRE are flipped to induce the dipole current for radiation in (2), an electric field $E$ is produced. This electric field is inversely proportional to the relative permittivity as shown in equation (7):

$$E \propto \frac{\sigma_q}{\varepsilon_r \varepsilon_o} \quad (7)$$

where $\varepsilon_r$ is the relative permittivity of the piezoelectric material. the radiated field strength for an antenna is determined by the maximum achievable current and its distribution. In the case of ADMIREs, the maximum current limit is determined by the charge density that results in electric near-fields just below the breakdown limit of the surrounding environment. Therefore, the maximum radiated field strength is directly proportional to the relative permittivity of ADMIREs. Fig. 2a compares a few commonly used piezoelectric materials with different values of relative permittivity. The same charge density of 1 mC/m² is assumed on the top and bottom surfaces while the generated electric field and the corresponding surface potential are calculated for a piezoelectric material with a thickness of 1 cm. Fig. 2b shows a piezoelectric material at resonance surrounded by air and its corresponding voltage distribution, where the fringing electric field is represented by the black arrows. For materials such as Quartz, AlN, LiTaO$_3$, and LiNbO$_3$ with low/moderate relative permittivity, the electric field is higher than or very close to the air breakdown field (~3MV/m), thus imposing a fundamental limit on the maximum radiation. Although one conceivable solution to this problem is non-metallic vacuum packaging, it increases both the antenna volume and cost, making it bulky and less reliable. On the other hand, an ADMIRE with a high relative permittivity such as PZT or PMN-PT ($\varepsilon_r > 1000$) can be used to mitigate this issue. In addition to enabling greater maximum radiation, better near-field confinement inside high permittivity piezoelectrics results in the near-field region becomes shortened to a fraction of the distance compared to the near-field of an equivalent infinitesimal electric dipole[20]. To facilitate future material selection for optimal antenna performance, the following figure of merit for ADMIREs is defined:

$$FoM = d^2 C^E \varepsilon_r Q \tag{8}$$

Orders of magnitude further improvement in radiation for acoustically driven antennas is expected with further optimization of material choice and design.

In addition to efficient radiation, passband transmission requires a modulation technique to send information. Simple and common digital modulation schemes can be utilized such as binary amplitude, frequency, and phase-shift keying (BASK, BFSK, BPSK) to directly modulate the ADMIRE (carrier) amplitude, frequency or phase with a modulating bit stream[23]. A mechanical antenna such as the ADMIRE has a settling time that is directly proportional to its quality factor and limits the BASK (on-off keying) rate since the mechanical system must be switched on and off corresponding to bit 1 and bit 0, respectively. The same applies to BPSK due to the phase discontinuity that requires the system to resettle and synchronize with the driving signal every time the phase changes. This presents a tradeoff between the material quality factor, which is required for efficient antenna operation, and the maximum achievable data rate, which is required for bandwidth efficiency[23]. On the other hand, BFSK can be designed to have a fixed amplitude and continuous phase, sometimes referred to as continuous phase FSK (CPFSK), or minimum-shift keying (MSK), which mitigates the amplitude settling limitation and dramatically improves the achievable data rates. An FoM presenting the characteristics of a BFSK modulator can be expressed as follows:

$$FoM_{Mod} = \Delta f \times FSK_{Rate} \tag{9}$$

where $\Delta f$ is the separation between the two frequencies representing the binary message ($\Delta f = f_2 - f_1$) and $FSK_{Rate}$ is the maximum achievable FSK rate for switching between the two frequencies. For practical systems, $\Delta f$ must be as large as possible to allow for larger separation between the band-pass filters (BPF) in the receiver, which relaxes the BPF design specifications and reduces the bit error rate (BER), while higher $FSK_{Rate}$ enables higher bit rates (for BFSK $Bit_{Rate} = 2 \times FSK_{Rate}$).

**Design**

Depending on the design goals, different resonance modes and frequencies can be targeted based on the piezoelectric material properties, dimensions, vibration direction, and excitation to meet performance metrics. In this paper, a high FoM ADMIRE antenna is designed to operate at the upper bound of the VLF band. Emphasis is placed on measuring the ADMIRE far-field radiation in the VLF band, and thus the FoM is constrained by frequency and geometry considerations and well below the ultimate FoM achievable for the ADMIRE. A disc resonator is designed with a high quality factor as shown in Fig. 3a and 3b, with 0.5 cm wide silver electrodes patterned on the sides of the PZT disc. This structure forms an acoustic resonator that is mechanically free with metal electrodes to drive it into resonance via the $d_{31}$ coefficient. The lateral vibration of the disc, also known as contour mode or dilation mode, is excited by applying a time-varying voltage on the metalized edges of the PZT disc. Upon excitation, the time-varying electric field introduced by the electrodes (configured as a pair of top and bottom electrodes) excites the piezoelectric disc into vibration via the reverse piezoelectric effect. The excited acoustic wave is reflected by the PZT disc boundaries, resulting in a standing acoustic wave with its maximum stress at the disc center. Fig. 3c shows the resonance dilation mode at 33.6 kHz along with the stress distribution.

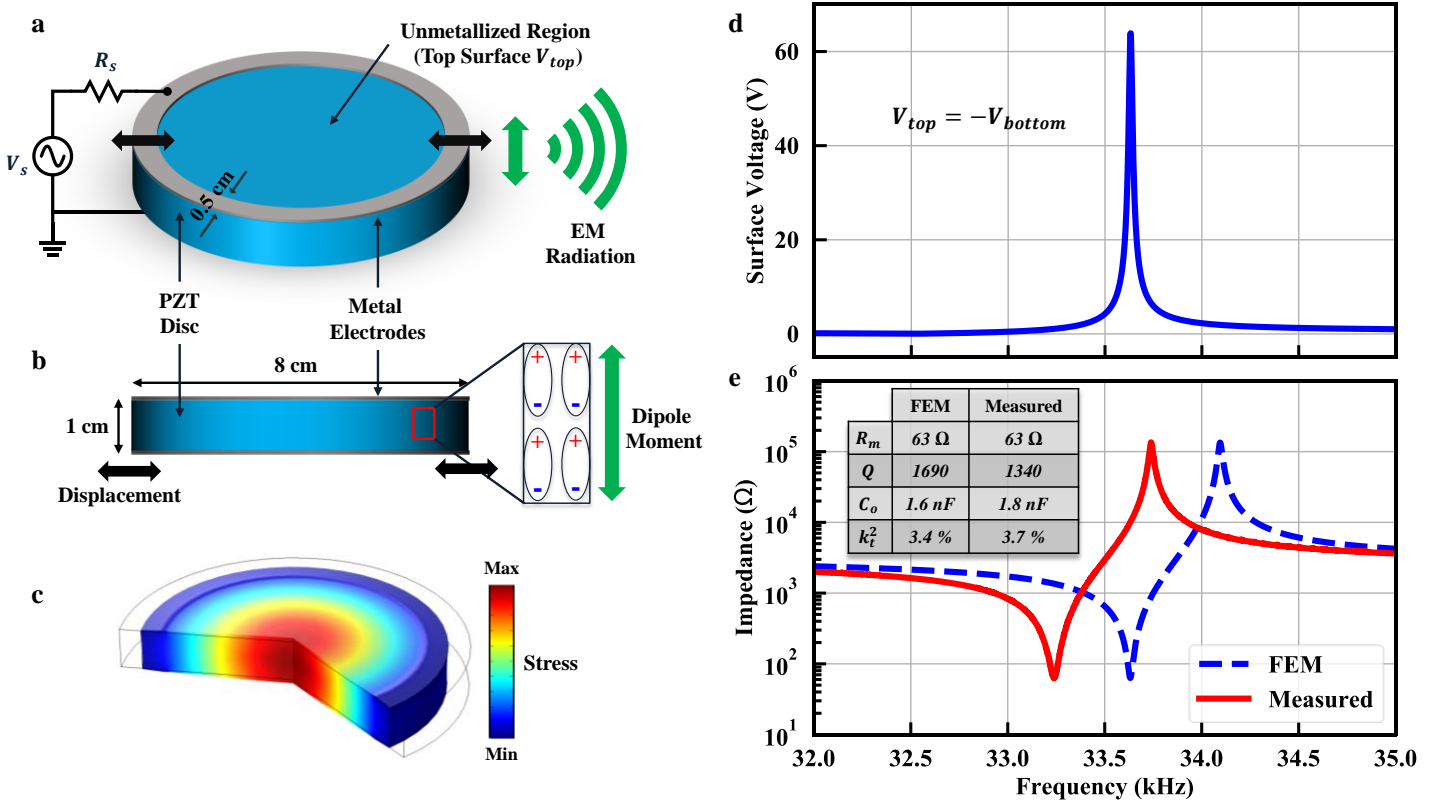

**Fig. 3 Proposed design and simulation of the piezoelectric radiating element using PZT. a,b**, The 3D view and the side view of the PZT disc (8 cm diameter, 1 cm thickness), with patterned silver electrodes along the disc circumference (0.5 cm width), driving source connected to the top and bottom electrodes, and dipole moment direction elaborated. **c**, Resonance mode showing the stress distribution at resonance formed by the acoustic standing wave. **d**, Simulated surface voltage with an applied voltage amplitude of 1 V. **e**, Simulated and measured impedances at the input terminals of the PZT disc.

During vibration, the mechanical stress generates electrical charges via the direct piezoelectric effect. The charges generated in the metalized electrode areas are neutralized by the electrodes, so the electrodes are designed around the edge of the disc where stress is lowest, leaving the highest stress, highest charge density center of the disc free to radiate. The density of the electrical charge is amplified by the quality factor at resonance, leading to a large time-varying dipole moment (current) that causes EM radiation. Additional geometries can be used to excite different high coupling piezoelectric materials in optimal resonant modes (such as dilation, thickness extensional or shear) to maximize generated charge, and thus radiation, due to higher piezoelectric coupling coefficients.

The structure is simulated using finite element modeling (FEM). Fig. 3d shows the generated surface voltage due to charge accumulation at resonance with an applied voltage amplitude of 1 V. A motional resistance of 63 Ω is designed to match with typical 50 Ω systems at the 33.6 kHz resonance, as seen in Fig. 3e which shows the impedance at the input terminals of the PZT disc (both simulated and measured). The motional resistance can be further tailored for perfect matching with 50 Ω systems by changing the width of the electrodes, since $R_{rad}$ is negligible for matching consideration ($R_{rad}/R_m \ll 1$). According to the BVD model $R_m$ can be expressed as:

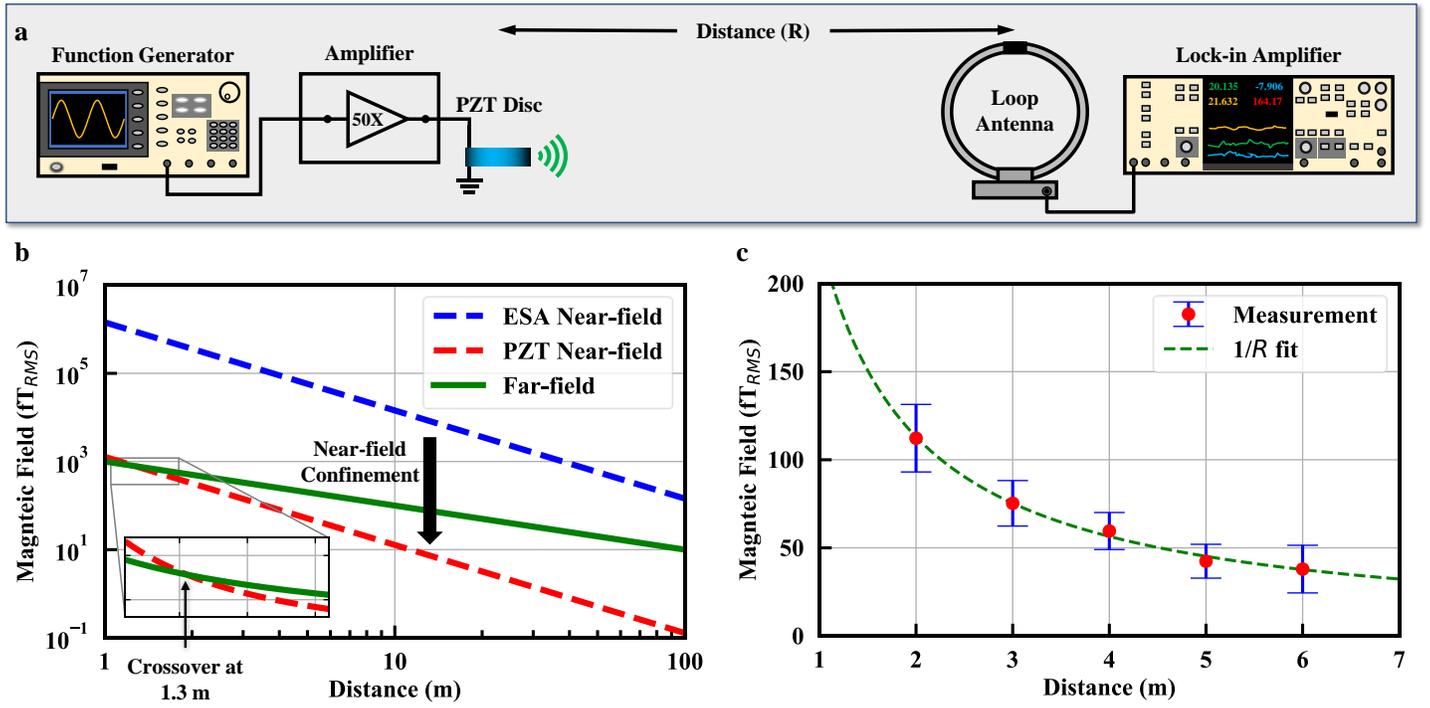

**Fig. 4 Simulation and measurement of the PZT radiation. a**, Measurement setup for detecting the magnetic field radiation of the ADMIRE. **b**, Simulation comparing ESA (infinitesimal dipole) and PZT magnetic fields. Due to the high relative permittivity of PZT, the magnetic field is confined within it, which dramatically reduces the near-field component relative to the equivalent ESA approximation. The radiated far-field can be measured very close to PZT after passing the crossover point at 1.3 m. **c**, Measured magnetic field vs. distance exhibiting clearly the far-field regime.

$$R_m = \frac{\pi^2}{8\omega_r C_o k_t^2 Q} \tag{10}$$

where $k_t^2$ is the electromechanical coupling coefficient.

## Results

To demonstrate the ADMIRE antennas, a prototype is created from a 1 cm thick, 8 cm diameter disc of PZT. A 20 µm thick, 0.5 cm wide silver ring electrode is patterned onto the top and bottom surfaces and driven to excite the PZT in the dilation mode via the $d_{31}$ piezoelectric coefficient. The resonant response is extracted from a direct impedance measurement and yields the results shown in Fig. 3e. The radiation measurements of the ADMIRE are complicated by the near-field confinement due to the high permittivity of the PZT. Unlike the far-field radiation of the ADMIRE which is dependent only on the equivalent current caused by the flipping dipole moments, the radiated near-fields are confined by the large relative permittivity of PZT within the dielectric. This means that near-field radiation, characterized by $1/r^3$ for electric fields and $1/r^2$ for magnetic fields, is diminished in both magnitude and distance. Compared to an equivalently sized 33 kHz ESA which radiates in the near-field regime up to 1 km, the ADMIRE antenna reaches its far-field regime (magnetic) after around 1.3 meters. Due to the respective distance scaling of $1/r^2$ vs. $1/r$, equivalent magnetic field radiation from the ESA is 100 times larger at ten meters than the ADMIRE radiating the same power. Therefore, both the PZT disc and the measurement setup shown in Fig. 4a are designed to minimize RF interference from the leads and connections so that the PZT radiation is not obscured.

The magnetic field vs. distance is measured in free space to minimize RFI and scattering using the setup shown in Fig. 4a. As seen in Fig. 4c, the measured magnetic field decreases as $1/r$ as expected from the simulations in Fig. 4b, confirming the PZT ADMIRE exhibits far-field radiation very close to the antenna. An input power of 1.2 W is supplied to excite the PZT disc. Radiation is measured using a passive loop antenna and the magnetic field is extracted from a lock-in amplifier measurement using the measured antenna factor $AF = B_{RMS}/\mu_o V_{RMS}$, where $B_{RMS}$ is the root mean square (RMS) magnetic field, $V_{RMS}$ is the voltage measured with the lock-in amplifier, and $\mu_o$ is the free space permeability. In order to better distinguish the measured radiation from noise, an average field reading is collected over two minutes at each distance. Extrapolating the measured data to 1 km yields a magnetic field of 0.23 fT$_{RMS}$ with an ADMIRE driving power of 1.2 W compared to a simulated magnetic-field of 0.5 fT$_{RMS}$. The discrepancy between the simulated and measured field strengths is likely due to imperfect earth ground effects[14], shifts in resonance due to ambient temperature changes, and effects from nearby radiators and reflectors.

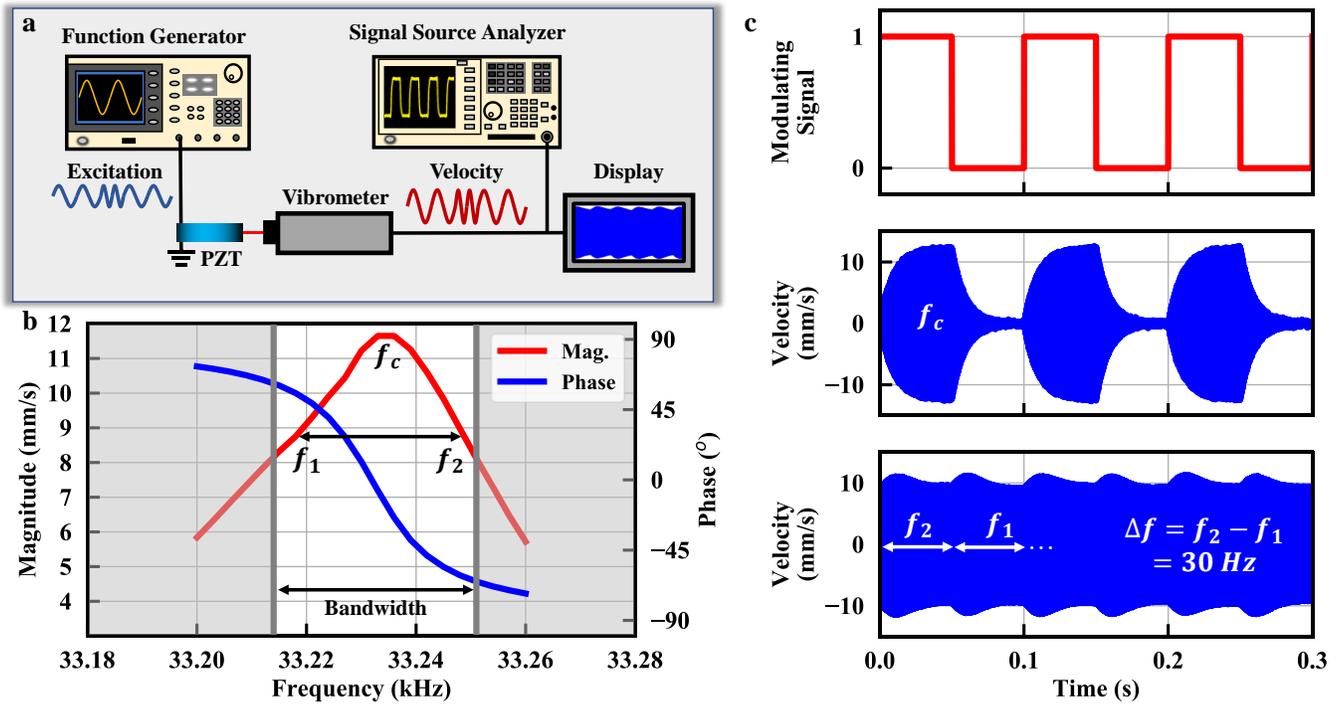

**Fig. 5 Different modulation schemes applied to the ADMIRE. a,** Mock-up of the measurement setup. **b,** Frequency response of the velocity (both magnitude and phase are measured) at the PZT disc edge. **c,** (top) 10 Hz bit stream, (middle) vibrometer velocity measurement of the amplitude-shift keyed (ASK) signal, and (bottom) vibrometer velocity measurement of the frequency-shift keyed (FSK) signal.

The PZT disc is directly modulated using a function generator outputting both ASK and FSK signals with the resonant response of the PZT disc captured using an optical vibrometer as shown in Fig. 5a. In both cases the 10 Hz binary bit stream at the top of Fig. 5c is used. With the ASK signal, as the driving signal is switched on and off the resonator energy ramps up and down over a duration inversely proportional to the loaded quality factor ($Q_L \approx 850$) The ramping time limits the fundamental modulation rate for direct BASK to approximately $1/2T$, where the time constant $T = 3 \times (2Q_L/\omega) \approx 24.4$ ms (corresponding to 95% settling from the peak value) for the demonstrated measurement. A fundamental design tradeoff must be considered to balance the inversely proportional data rate with the high $Q$ desired for the radiation FoM. BFSK modulation is conducted within the 3-dB bandwidth of the PZT resonator corresponding to BFSK frequencies of $f_1 = 33.218$ kHz and $f_2 = 33.248$ kHz. The input is a continuous-phase FSK with no discontinuity when switching between the two frequencies. However, due to the phase difference of the mechanical resonator at the two frequencies, the mechanical resonance is out of phase with the modulated driving signal when it is switched and ramping of the PZT edge velocity occurs while energy is transferred from one resonant frequency to another as seen in Fig. 5c(bottom). As the modulation frequency approaches the limit set by the frequency settling, although the amplitude of resonance is not diminished, the demodulated output signal is distorted as seen in Fig. 6a and 6b. Multiple approaches can be implemented to surpass the $Q$-limited fundamental modulation rate of the resonator by ensuring that the phase of the resonator and driving signal are in phase during modulation transitions.

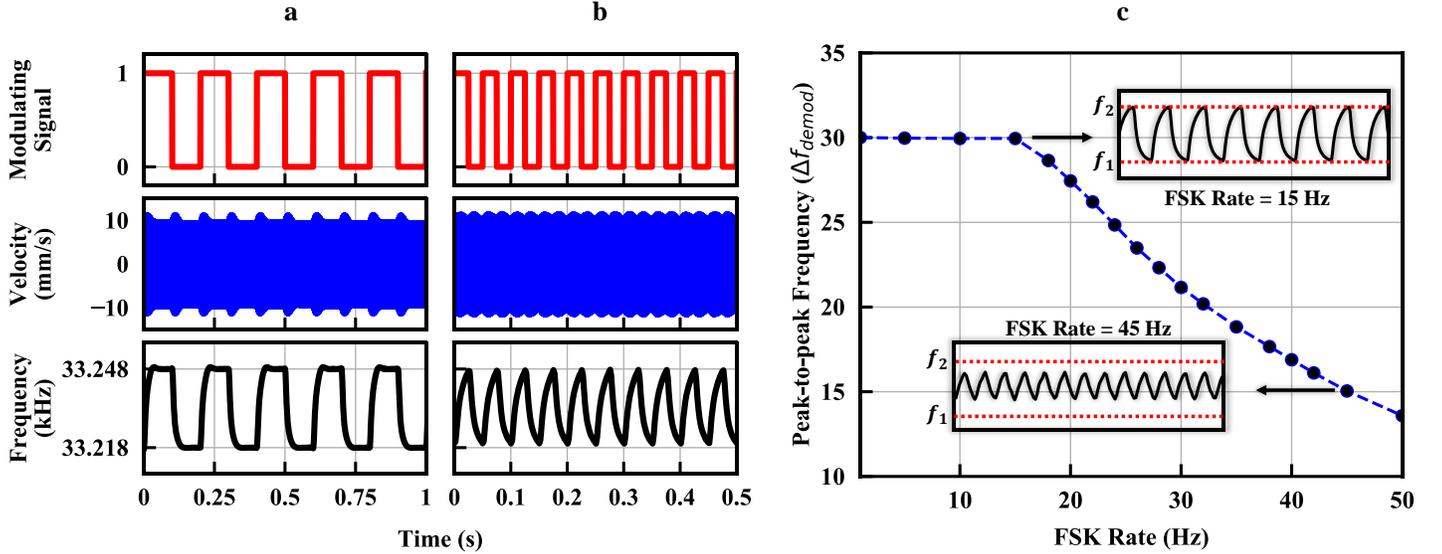

**Fig. 6 Measured FSK modulation of the PZT ADMIRE at different FSK data rates. a,b,** 5 Hz**,** and 20 Hz FSK rate. **a,b,** (Top) modulating signal, (middle) measured velocity, and (bottom) demodulated signal using signal source analyzer (SSA). **c,** Peak-to-peak frequency difference ($\Delta f_{demod} = f_2 - f_1$) after demodulation with SSA at different FSK rates. The upper limit for direct modulation using the 3-dB bandwidth is limited by frequency settling, the maximum direct modulation frequency is approached resulting in a distorted modulation waveform.

## Conclusion

The presented ADMIRE demonstrates the potential for portable VLF transmitters that have been unattainable for decades. Although significant work is needed to optimize both antenna efficiency and bit rate, this work provides a framework for future development in the field of acoustically driven antennas.

## Methods

### Modeling

Piezoelectricity and EM radiation modeling require multidisciplinary understanding and coupling between the electrical and the mechanical domains. This is achieved by using FEM available from "COMSOL Multiphysics" that couples these domains in the "piezoelectric devices" toolbox. Such a model can be used to determine the resonance frequency of the structure using eigenfrequency simulations followed by frequency domain simulations to find out parameters such as induced stress/strain, the velocity of the edge of the disc, internal polarization, surface voltages, and admittance. The internal polarization (charge density) can then be used to calculate the polarization current and the radiated EM field from equations (2) and (3). PZT piezoelectric properties are supplied by the vendor (see Supplemental Information) and input to the FEM model. The quality factor is modified so that the motional resistance $R_m$, calculated using equation (10), matches the measured value. The simulation time can be dramatically reduced since our designed PZT disc exhibits symmetry around its central axis so, axisymmetric simulations are utilized. In addition, an air sphere is added to model the surrounding of the ADMIRE which enables near-field simulations around the ADMIRE to compare air breakdown around different piezoelectric materials. Simulated admittance is compared to measured admittance in Fig. 3e. Moreover, the simulated near-field of PZT is compared with the infinitesimal dipole near-field in Fig. 4b.

### PZT Fabrication and Characterization

The PZT discs are commercially fabricated by Physik Instrumente (PI) and made from their PIC181 material[24]. The commercial discs have an 8 cm diameter and a 1 cm thickness with both top and bottom surfaces fully metalized with ~20 µm of silver. Patterning of the silver is conducted using an end mill to remove the interior metal until only the desired 0.5 cm ring along the edge remains. Two wire leads are split from a BNC cable and soldered to the top and bottom metal surfaces to provide electrical excitation, with the lead lengths minimized to reduce near-field radiation from the current loop.

Characterization of the PZT is conducted by connecting a Tektronix AFG3152C function generator directly to the electrodes via BNC cable. An Agilent E4445A spectrum analyzer connected in series with the PZT disc is then used to characterize the impedance of the PZT as a function of frequency, from which the motional resistance, electromechanical coupling ($k_t^2$) and mechanical quality factor are extracted. The bottom surface of the PZT disc rests freely on a 2x2 cm insulating cardboard lattice and the top and side surface are unconstrained. Multiple clamping configurations were considered but yielded negligible changes in mechanical properties. Input power to the PZT disc is characterized by removing the series spectrum analyzer and adding an Agilent MSO7104B oscilloscope in parallel with the disc with the power dissipation measured from the voltage drop across the PZT.

**Radiation Measurement**

Wireless radiation measurements of the generated magnetic field are conducted in an open environment to minimize scattering and noise. Confinement of the near-field component of the PZT radiation results in the far-field component dominating beyond 2 m but current loops in the transmitter exhibit near-field dominate radiation up to 1000 m. In order to minimize near-field radiation from current loops, leads and connections are minimized and oriented to exhibit radiation orthogonal to the receiving antenna. The resulting total radiation exhibits a near-to-far-field crossover between 1 and 2 meters. At the operating frequency, only the PZT radiation can exhibit a $1/r$ roll off at distances less than 1 km, therefore all measured radiation with a $1/r$ fit is attributed solely to the PZT.

The transmitting system consists of a Tektronix AFG3152C function generator connected in series to a 50x Trek model 2100HF amplifier to generate a sufficiently large excitation to measure the far field. The amplifier presents a resistance of 200 Ω in series with the 63 Ω motional resistance of the PZT disc at resonance which results in a diminished loaded quality factor where $Q_L \approx QR_m/(R_m + R_s)$. From (4), the diminished $Q_L$ results in a lower radiation efficiency for the PZT and a higher power is needed to drive the loaded ADMIRE.

The magnetic field is measured using an AH-Systems SAS-565L 24" shielded passive loop antenna which is oriented to receive the maximum signal from the PZT far-field component. Incident radiation induces an open circuit voltage across the antenna terminals proportional to the field strength. The antenna factor of the loop receiver is calibrated by the manufacturer post-production to be 1.74 Ω$^{-1}$m$^{-1}$ at 33 kHz and is used to extract the measured B-field where $B = \mu_0 V * AF$. The open-circuit voltage of the antenna is measured using a Stanford Research Systems SR865A lock-in amplifier that is frequency locked to the transmitting PZT disc and employs a 24 dB/octave bandpass filter to attenuate noise around the locked frequency. Measurements were made at 1 m distance intervals for 2 minutes at a time using a 1 second time constant. The measured B-field strength is extracted from the average terminal voltage over the 2-minute measurement window, with 1 standard deviation error bars also provided to account for variance in the measured field strength due to noise. Between field measurements, the noise floor is measured at 1-minute intervals with the input signal turned off (Supplementary Fig. 3). Measurements beyond 6 m exhibit a signal-to-noise ratio < 2 and are not shown.

**Modulation Measurement**

Direct digital modulation of the ADMIRE can be done by altering amplitude, frequency, and phase of the excitation signal which in turn modulates the mechanical resonance of the ADMIRE and thus radiated signal. In this paper, we focus on BFSK since it has a continuous phase which lowers the mechanical settling time compared to both BPSK and BASK. The modulation is evaluated using the measurement setup in Fig. 5a. The setup consists of a Tektronix AFG3152C function generator that directly excites the ADMIRE with continuous phase BASK or BFSK signals. A Polytec OFV-5000 laser vibrometer is used to measure the velocity of the PZT edge while vibrating in the dilation mode. The two BFSK modulation frequencies are chosen to be within the 3-dB bandwidth as shown in Fig. 5b. Fig. 5c shows the modulated velocity of BASK and BFSK with a 10 Hz modulation rate. The velocity signal is then input to a ROHDE & SCHWARZ FSUP signal source analyzer with FM demodulation capability to demodulate the signal as shown in Fig. 6a and 6b (bottom figures) for 5 Hz and 20 Hz BFSK rates, respectively.